# Novel Muscle Monitoring by Radiomyography(RMG) and Application to Hand Gesture Recognition

Zijing Zhang* and Edwin C. Kan, *Senior Member, IEEE*

*Abstract*— Conventional electromyography (EMG) measures the continuous neural activity during muscle contraction, but lacks explicit quantification of the actual contraction. Mechanomyography (MMG) and accelerometers only measure body surface motion, while ultrasound, CT-scan and MRI are restricted to in-clinic snapshots. Here we propose a novel radiomyography (RMG) for continuous muscle actuation sensing that can be wearable and touchless, capturing both superficial and deep muscle groups. We verified RMG experimentally by a forearm wearable sensor for detailed hand gesture recognition. We first converted the radio sensing outputs to the time-frequency spectrogram, and then employed the vision transformer (ViT) deep learning network as the classification model, which can recognize 23 gestures with an average accuracy up to 99% on 8 subjects. By transfer learning, high adaptivity to user difference and sensor variation were achieved at an average accuracy up to 97%. We further demonstrated RMG to monitor eye and leg muscles and achieved high accuracy for eye movement and body postures tracking. RMG can be used with synchronous EMG to derive stimulation-actuation waveforms for many future applications in kinesiology, physiotherapy, rehabilitation, and human-machine interface.

## I. INTRODUCTION

### A. Muscle Monitoring

Monitoring skeletal muscle activities has significant medical and commercial applications, including detection of muscle fatigue and injury, diagnosis of neuromuscular disorders, assessment for physical training and rehabilitation [1]-[3], human-computer interface (HCI) [4], and robotic control [5]. Electromyography (EMG) is presently the prevalent muscle sensing technique, which measures the neural stimulation during continuous muscle contraction by intramuscular needle/wire electrodes or by epidermal electrodes on the bare skin [6]. However, EMG does not provide direct quantification of muscle contraction. Mechanomyography (MMG) and accelerometers record the mechanical motion on the body surface, but lacks information in deep muscle layers [7]. Ultrasound monitoring requires body probes with surface preparation for impedance match [8]. Magnetic resonance imaging (MRI) and computer tomography (CT)-scan can obtain high-resolution muscle imaging, but is expensive and immobile, providing only short snapshots of muscle motion in a dedicated clinical setup [9].

Hence, a direct muscle activity sensor that can accurately and continuously track muscle contraction in the superficial and deep layers with high user comfort is a critical complement to EMG in many biomedical and HCI applications.

### B. Hand gesture recognition by muscle actuation detection

Hand gestures are controlled by complex muscle groups, many of which are extended from the forearms. Wearable sensors on wrist or arm bands for hand gesture recognition (HGR) [10][11] is of high interest to facilitate HCI [12][13] and a myriad of other applications [14][15]. Current HGR techniques however have many limitations. Vision-based system requires off-body line-of-sight (LoS) cameras, and is vulnerable to self and ambient obstruction [16]-[18]. Depth-perception methods demand excessive geometry reconstruction computation [19][20]. Gloved-based sensors can hinder hand motion, and are often inconvenient and uncomfortable [21]. Accelerometers can only transduce surface motion, and are impractical and cumbersome if deployed on fingers or phalanges [22]. The HGR radar, such as Google Soli [23], has to be in the LoS path to the target hand and can also suffer from path noises [24][25].

By monitoring muscle activities on the forearm, an HGR system can be built with user convenience for many gestures, such as the conventional surface electromyography (sEMG) tracking the neural stimulation of forearm muscles [26]-[30]. However, sEMG is limited by the ambiguity in the skin potential, vital-sign interference, and requirement of numerous electrodes with high-quality contact to the bare skin, sometime raising health concerns in long-term wearing and suffering low user acceptance. Electrical impedance tomography (EIT) recovers the interior impedance geometry of arm muscles, but resolution is limited and cross-user generalization is often questionable [31].

### C. RMG for muscle activity sensing

In this paper, we propose radiomyography (RMG), a novel skeletal muscle sensor that can non-invasively and continuously capture muscle contraction in various layers. RMG uses

This work is supported by Department of Defense of United States through Office of the Congressionally Directed Medical Research Programs (CDMRP) Discovery Award PR-182496, and by National Institute of Health (NIH) R21 DA049566-01A1.

Z. Zhang, and E. C. Kan are with School of Electrical and Computer Engineering, Cornell University, Ithaca, NY 14853, USA. E-mails: {zz587, eck5}@cornell.edu.

multiple-input-multiple-output (MIMO) near-field coherent sensing (NCS) radio signals [32][33] to measure the dielectric property change and dielectric boundary movement of nearby muscle groups. NCS couples ultra-high frequency (UHF) electromagnetic (EM) waves inside the body and reads out the internal organ and tissue motion signals as modulated antenna characteristics or scattering impedance matrices [32]. As the near-field coupling is nonlinear and convoluted in the capture volume from different observation points, we explore spatially diverse channels to distinguish the detailed muscle action. MIMO provides $N^2$ observation channels in 3D from $N$ sensing units on the body surface to enhance this critical spatial diversity [33]. Our experimental RMG prototype employs 4 sensing antenna pairs for 16 channels in total, and can be extended to much larger number of channels with modest system cost.

Radio-frequency (RF) signals in the UHF band, especially in the near-field region, will penetrate most dielectrics effectively without requiring direct skin contact. Therefore, our RMG system can be wearable over clothing or installed in a nearby off-body apparatus such as armrests and wrist pads.

To demonstrate that RMG can monitor the superficial and deep muscles, we carried out continuous recording of the complex forearm muscle contraction during various hand gestures by the MIMO channels, which provide rich information to decode the convoluted muscle activities. To validate HGR by RMG, we performed a human study of 8 participants with 23 gestures, including motion of fingers, palm, and wrist in different speeds and multiple degrees of freedom (DoFs). Various sensor modalities and forearm positions were also tested. After segmenting data from the RMG sensor, we transformed the 1D time waveforms to 2D time-frequency spectrograms using the short-time Fourier transform (STFT) and continuous wavelet transform (CWT). For gesture classification, we adopted vision transformer (ViT) as the deep-learning model [34] to compare with conventional convolutional neural networks (CNN). To provide a baseline comparison, we also benchmarked RMG with simultaneous sEMG for correlation and timing verification. To broaden the application scope, we also deployed leg RMG and radiooculogram (ROG), which tracked leg and eye muscle activities, respectively.

RMG can be applied to numerous applications. Gesture recognition and eye movement detection can be used as a middleware for HCI, such as virtual reality (VR) control and cybersickness reduction. In clinical applications, RMG can be used as assessment for voluntary and evoked muscle contraction, disorder of muscle functions, and rehabilitation. RMG can be integrated with EMG for possible diagnosis of neuromuscular disorders including Parkinson's disease, as well as for feedback control of electromyostimulation (EMS). ROG can be further applied to rapid eye movement (REM) monitoring with eyes closed during sleep.

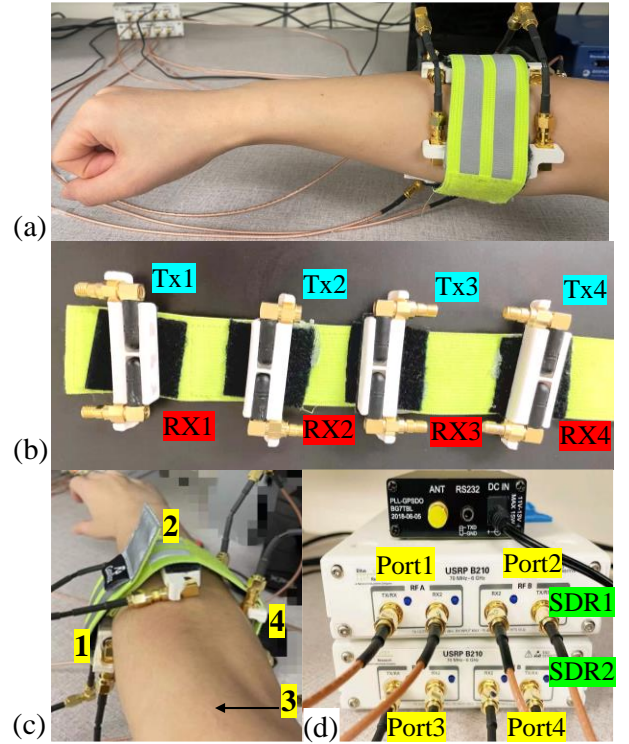

**Fig. 1.** (a) Forearm placement of a wearable RMG by SDR. (b) Four sensing antenna pairs attached to the armband. (c) The cross section view of RMG placement of 4 measuring points. The third probe is on the anterior side of the forearm. (d) The transceiver s setup by two software-defined radios.

## II. OPERATIONAL PRINCIPLES

*B. Coupling into the deep muscle groups*

Muscles in a forearm are divided into anterior and posterior muscles, both containing superficial and deep layers. Hand gestures by the superficial muscle layers can be captured by motion sensors like accelerometers with tight skin contact. However, deep-layer muscles are critical for HGR but can raise ambiguity for surface-based sensors. The forearm muscles actuating the hand gestures are listed in Table 1. For example, deep posterior muscles are important for thumb and index finger movements. Hence, forearm muscle sensors for hand and wrist gestures will be able to differentiate hand gestures reliably only if all muscle groups, not just the surface ones, are included in the sensor readout.

flexor digitorum profundus is the only major muscle that can flex the distal interphalangeal joints of the fingers, and four of RMG provides a new solution to detect muscle actuation for accurate HGR. From near-field EM coupling, the muscle motion in various layers is directly modulated onto multiplexed radio signals. Previous radar-based systems often operated in the far field when the EM energy is mostly reflected at the body surface, so only the surface motion would be captured [35]. In comparison, NCS has been validated for deep coupling into human body to monitor heart valve motion [33][36], femoral pulses [37], and diaphragmatic breathing [38][39].

TABLE 1. MAJOR MUSCLE GROUPS GENERATING THE HAND GESTURES

| Basic Gesture | Step 1 | Major Muscles | Step 2 | Major Muscles |
|---|---|---|---|---|
| Grasp | Extend 5 fingers | ED, EPL, EDM | Flex 5 fingers | FDP, FDS, FPL |
| Point Thumb | Extend thumb | EPL | Flex thumb | FPL |
| Point Index | Extend index | EI | Flex index | FDP, FDS |
| Point Ind.+Mid. | Extend ind.+mid. | EI, ED | Flex ind.+mid. | FDP, FDS |
| Point 4 Fingers | Extend 4 fingers | EI, ED, EDM, | Flex 4 fingers | FDP, FDS |
| Fist | Flex 5 fingers | FDP, FDS, FPL | Rest | |
| Wrist Up | Extend wrist | ECU, ECRL, ECRB | Flex wrist | FCU, FCR |
| Wrist Down | Flex wrist | FCU, FCR | Extend wrist | ECU, ECRL, ECRB |
| Muscle groups: ECU: Extensor Carpi Ulnaris; ECRL: Extensor Carpi Radialis Longus; ECRB: Extensor Carpi Radialis Brevis; FCU: Flexor Carpi Ulnaris; FCR: Flexor Carpi Radialis; ED: Extensor digitorum; EDM: Extensor digiti minimi; FDS: Flexor Digitorum Superficialis; EPL: Extensor Pollicis Longus; EI: Extensor Indicis; FDP: Flexor Digitorum Profundus; FPL: Flexor Pollicis Longus. Green font: Superficial; Blue font: Intermediate; Red font: Deep. ||||| 

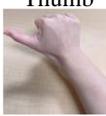
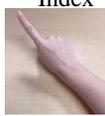
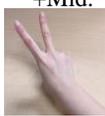
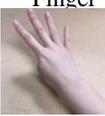
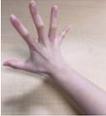
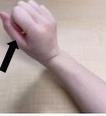
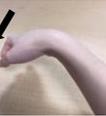
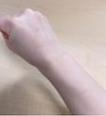
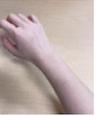

| | Finger Gestures | | | | | Wrist Gestures | | Other | |
|---|---|---|---|---|---|---|---|---|---|
| Basic Gesture | Point Thumb | Point Index | Point Ind.+Mid. | Point 4 Finger | Grasp | Wrist Up | Wrist Down | Fist | Rest |
| Quick | P1 | P2 | P23 | P4 | G | U | D | | |
| Quick Double | P1×2 | P2×2 | P23×2 | P4×2 | G×2 | U×2 | D×2 | | |
| Slow | sP1 | sP2 | sP23 | sP4 | sG | sU | sD | sF | R |

**Fig. 2.** 23 hand gestures used in the study protocol.

*B. MIMO: Rich $N^2$ channel by N points*

In this work, we adopt MIMO to incorporate $N^2$ usable channels from $N$ observation points [33]. MIMO is a mature RF technique where different transmitters (Tx) can be well isolated by either frequency or code multiplexing. Similar techniques can be employed by colors in vision and subcarriers in ultrasound, but RF MIMO offers higher channel isolation than optical or acoustic waves with much lower cost thanks to the mature wireless industry. $N$ Tx can then be simultaneously received and demodulated by $N$ receivers (Rx) to accomplish $N^2$ synchronous channels to fulfill the spatial diversity requirement to observe complex 3D geometry and motion. Due to tissue dispersion and near-field nonlinearity, the channel by Tx1–Rx2 would represent different information from Tx2–Rx1. Our RMG prototype utilized $N = 4$ sensing points around the forearm, and collected signals from 16 channels.

## III. EXPERIMENTS

*A. Experimental setup*

The four pairs of the RMG sensing antennas were attached to a wearable armband on the middle forearm as shown in Fig. 1(a). Each sensing unit consisted of two monopole whip antennas (Taoglas TG.19.0112) mounted on a 3D-printed holder. The overall dimensions for each unit were $69 \times 17 \times 11$ mm. As shown in Fig. 1(b), each sensing unit has a parallel orientation to the forearm muscle for enhanced coupling to the muscle contraction or relaxation. Unit 1 was placed close to extensor pollicis longus and flexor pollicis longus, which controlled extension and flexion of the thumb. Unit 2 was placed close to the extensor muscles, which produced extension at the wrist and fingers. Unit 3 was placed close to the flexion muscles, which were associated with pronation of the forearm, as well as flexion of the wrist and fingers. Unit 4 was close to flexor digitorum profundus which flexed the four fingers except the thumb. Multi-channel observation can help decode the convoluted muscle motion in various hand gestures by sensor proximity and rich MIMO channels.

The RMG transceiver was implemented by two synchronized software defined radios (SDR, Ettus B210). The SDRs were connected to a host computer through universal serial bus (USB), and the control software was implemented in LabVIEW. Each port of the MIMO system consisted of one Tx and one Rx, which was then connected to one sensing antenna pair. Each SDR supported two synchronous ports. Note that the present RMG on the armband were connected by cables to off-body SDR for flexible prototyping of RF transceiver parameters. An all-in-one wireless unit of RMG can be a straightforward extension in the future [40].

The digital baseband tone $f_{BB}$ of each Tx went through the digital-to-analog converter (DAC) and was then mixed with the carrier frequency $f_{RF}$ in a standard quadrature scheme. The RF power was less than −10 dBm or 0.1 mW, well under the safety limits set by occupational safety and health administration (OSHA) in the UHF band [41]. The Tx signal was coupled into the forearm muscle groups, received by all Rx, and then demodulated and sampled by the analog-to-digital converter (ADC) to retrieve the baseband. $f_{RF}$ was selected at 900 MHz for deep muscle penetration and multiple Tx channels utilized frequency-division multiple access (FDMA) for $f_{BB}$. We configured the dual SDR as 4 self-channels and 12 cross channels in the MIMO setup.

*B. Human study protocol*

RMG was tested on 8 healthy participants (Supplementary Table 1). We designed 23 gestures including finger, palm, and wrist motions with various speeds and multiple DOFs as shown in Fig. 2. We had 8 dynamic basic gestures and 1 static resting gesture. Basic gestures were extended to three versions including quick, double quick, and slow, except that the gesture 'Fist' only had the slow version. These gestures are chosen for their common rendition in HGR, as well as for confirmation of deep muscle sensing. Every gesture was performed in a fixed time window of 5 s. All gestures excluding 'Rest' were dynamic and comprised two steps as described in Table 1. For the quick version, step 2 was performed immediately after step 1, while the slow version had a holding time around 2 s between steps 1 and 2. For each dynamic gesture, after step 2, the hand would relax back to the 'Rest' gesture (Supplementary video).

## IV. DATA PROCESSING

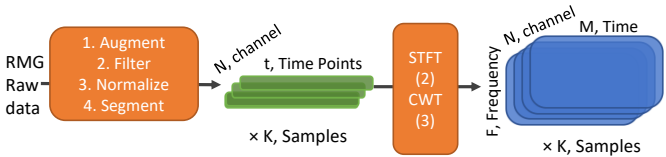

**Fig. 3.** Schematic for data pre-processing to feed spectrogram into machine learning.

After collecting data on multiple participants, we processed the raw data before feeding the output to the machine learning (ML) models for classification. The signal processing before learning help de-noise the data set and avoid overfitting, which will become apparent when comparison was made against the ML model based on raw waveforms. The data flow is shown in Fig. 3.

*A. Multi-channel augmentation*

From the MIMO configuration in RMG, we obtained 16 channels on a forearm. Each channel contained the baseband phase $NCS_{ph}(t)$ and amplitude $NCS_{am}(t)$ in the quadrature scheme on $f_{BB}$. In addition to employing phase and amplitude, we also augmented the original complex number as part of the information to retain the intricate relation between $NCS_{am}(t)$ and $NCS_{ph}(t)$. Therefore, for 16 physical channels, we had 48 temporal series in total.

*B. Filtering and segmentation*

The 48 1D waveforms in time was then processed by:
1. Bandpass filtering (0.1 Hz to 5 Hz).
2. Waveform normalization with center 0 and standard deviation 1.
3. Waveform segmentation into individual windows of $T_{win}$ = 5 s. Each window now contained one gesture guided by voice instruction.
4. Annotation of the instructed gesture for each window.
5. Waveform detrending by subtracting the best-fit linear line from the data.

*C. 1D waveforms to 2D spectrograms*

In this work, we employed STFT and CWT to generate 2D spectrograms before feeding into the machine learning (ML) model. Transformation of 1D time waveforms to 2D time-frequency spectrogram would bring forth significant improvement in accuracy. We explored 2 STFT outputs with different window lengths and 3 CWT outputs with different mother wavelets (Supplementary Fig. 1). The ensemble of five 2D spectrograms from different perspectives were incorporated into ML for classification.

## V. RESULTS AND ANALYSES

*A. Classification by vision transformer*

The final output dataset from all 8 subjects consisted of 5,847 samples of 23 gestures to be fed into the ML model for HGR. Though classical ML models can be computationally less expensive, algorithmic and hardware improvements in recent years have facilitated complex neural networks on embedded systems efficiently [42]. We implemented vision transformer (ViT) as the classification ML model.

ViT has a deep-learning architecture inherited from the transformer model in natural language processing (NLP) [43] and is now gaining popularity in computer vision. To benchmark ViT performance, we also built a conventional CNN classifier.

Over the ViT architecture, patches of the 2D input image were constructed from the time-frequency spectrogram. The image was split into fixed-size patches (size = 5), each of which are then linearly embedded (dimension = 512). Position embedding was added, and the resulting vector sequences were fed to a standard transformer encoder. Inside the encoder, we had 6 transformer blocks, 16 heads in the multi-head attention layer, 64 dimensions of the multi-layer perceptron (MLP) (feed-forward) layer, and the dropout rate was set to 0.1. In CNN, we convolution layers, each followed by a BatchNorm layer, and 2 linear layers came after the convolution layers. Adam optimizer is used for both ViT and CNN.

*B. The personal training model*

We evaluated the classification accuracy of RMG by different cross-validation (CV) methods, feature sets and deep learning models. First, we built the personal training model for each participant. From individual person's dataset, each gesture was repeated around 30 times, and the total sample number was around 700 – 800. K-fold ($k$ = 7) CV was performed to estimate the mean accuracy for each participant. An overall accuracy was averaged on results from all participants. Fig. 4(a) shows the overall confusion matrix of ViT by the personal training model, which is normalized to the percentage of samples. RMG achieved an overall accuracy of 99.0% for 23 gestures in total, which employed the ensemble method by majority voting of all feature sets from 2 STFT and 3 CWT versions. We also compared the results using different feature sets separately and present the results in Fig. 4(b). The ensemble method with the flexibility to choose among all alternatives achieved the highest accuracy.

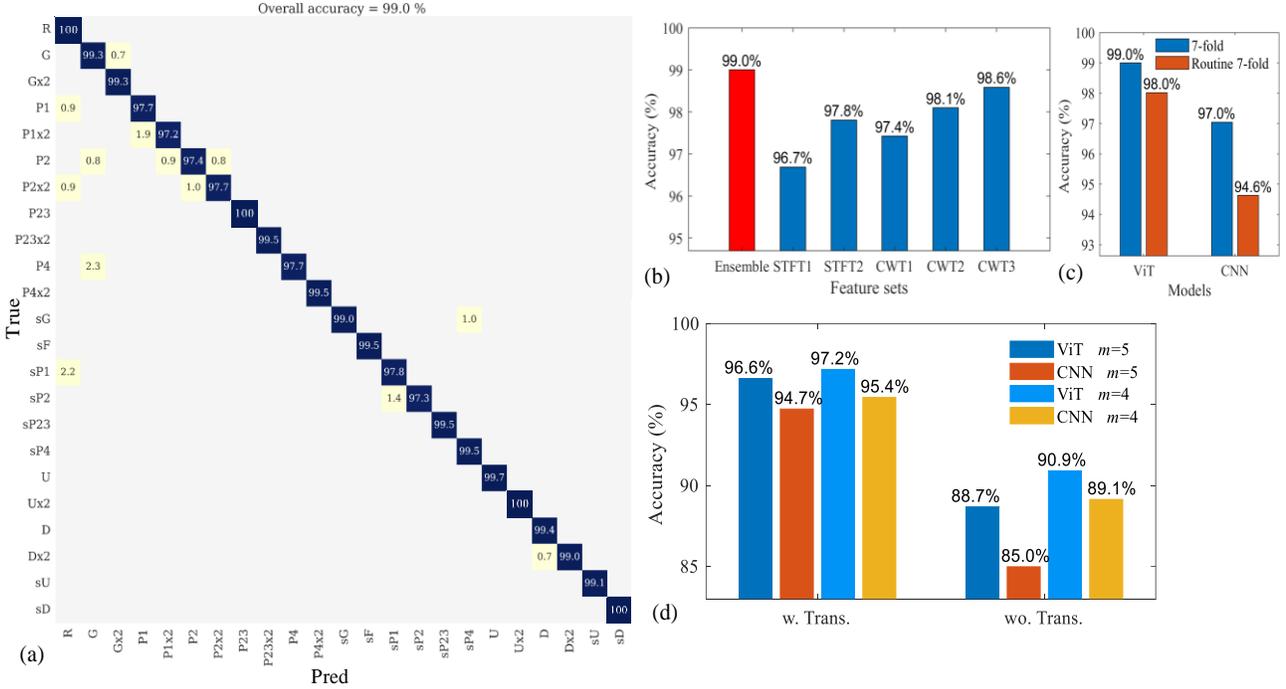

**Fig. 4.** (a) The confusion matrices showing the overall accuracy on all participants using the personal training model. (b)(c) Performance of the personal training model. (b) Accuracy using different transforms; (c) Accuracy by different ML models. (d) Accuracy with and without transfer learning for *m* = 4 or 5 by ViT and CNN on new participants.

Data were collected based on 5-min routines (around 60 gesture repetitions) in the hour-long testing for each participant to allow some rest and adjustment. Considering the hardware restart and sensor position alteration in each routine reset, signals collected by different routines can have more variations. Therefore, we presented another CV process, where all routines were independently tested, i.e., the gestures in the same routine were never divided between training and testing. As shown in Supplementary Fig. 2(a), routine-independent CV still achieved a high accuracy of 98.0%, which validated the system robustness against the hardware reboot and small sensor position variation in practical applications.

ViT was compared to CNN in Fig. 4(c), where the accuracy dropped from 99.0% in ViT to 97.0% in CNN for personal training CV, and from 98.0% in ViT to 94.6% in CNN for routine-independent CV. To illustrate the advantage of 2D spectrogram, we also built a 1D-CNN model using the time waveforms directly. Accuracies dropped from 97.0% in 2D CNN to 88.5% in 1D CNN for personal training CV, and similarly from 94.6% to 88.0% for routine-independent CV.

We adopted the MIMO setup for RMG in order to collect both self and cross channels. We also tested the recognition accuracy using only self-backscattering channels as the input for the ViT model. Accuracy degraded from 99.0% to 95.0% in the personal training CV. Similar degradation occurred when the number of antenna pairs decreased to 3. These showed the critical role of channel spatial diversity in HGR.

*C. Transfer learning for unseen users*

The HGR system must be robust to various practical conditions, especially for subject variation. Not only people perform hand gestures differently, but also the forearm size and muscle conditions have considerable distribution. Here, we adopted transfer learning (TL) [44] where we leveraged a pre-trained model with large amount of data from multiple users to test on a new user with a small amount of individual training data. To validate the TL effectiveness in RMG, we first generated the pre-trained model using all data from 7 participants. We then fine-tuned the model with *1/m* data from the new participant as short personal calibration. The final model was tested on the rest (*1 − 1/m*) data. This CV process is similar to *k*-fold, but only one fold is for training, and (*m − 1*) folds are for testing. Supplementary Fig. 2(b) shows the accuracies for all participants rotating as the new test case by the above TL strategy with *m* = 5. The averaged accuracy is 96.6%, and the normalized confusion matrix is presented in Supplementary Fig. 3.

ViT also outperformed CNN for our alternative TL strategy [34]. Fig. 4(d) shows accuracies for *m* = 4 or 5 with and without TL in the ViT and CNN models. ViT achieved higher accuracy than CNN in every scenario. Direct learning from *1/m* data without TL had much lower accuracy than the pre-trained model by TL. When the personal training set increased from ⅕ to ¼, accuracy also noticeably increased, indicating the trade-off between high accuracy and the amount of personal training data.

## D. Variations in experimental designs

TABLE II. EVALUATION FOR POSITION AND DESIGN VARIATIONS

|  | $d = 3$cm with TL | $d = 3$cm without TL | Notch | Box | Wrist |
|---|---|---|---|---|---|
| Accuracy (%) | 97.2 | 87.2 | 99.0 | 97.4 | 95.8 |

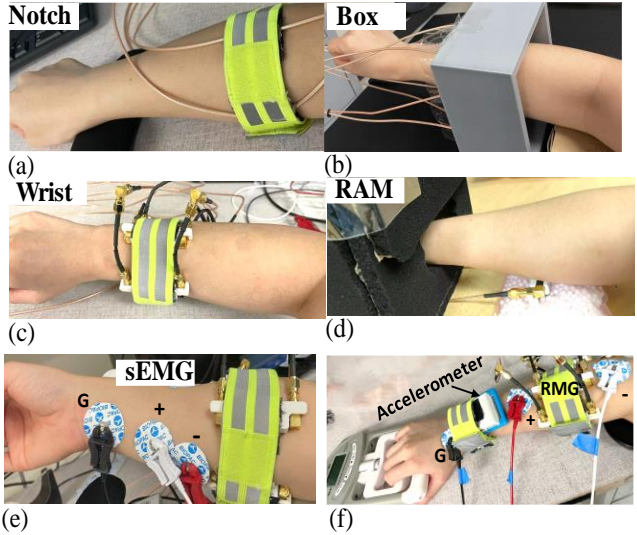

**Fig. 5.** Experimental setups of various designs: (a) A notch RMG; (b) A non-contact box; (c) A wristband; (d) Verification by the hand inside an RAM box; (e) Benchmark with sEMG with short +/− separation; (f) Slow grasp strength testing with sEMG and accelerometer.

Apart from subject dependence, accuracy degradation can also be induced from the sensor placement on the forearm. To test the adaptivity against sensor position variation, we performed another test on one participant with the same protocol but with the sensor position moved to a higher position by $d = 3$ cm. We used the same TL strategy to achieve the result in the first two columns of Table II. After adopting TL, accuracy was boosted from 87.2% to 97.2%.

For the human study above, we used the antenna-based sensing unit on the forearm. We also explored more sensor design variations. The first design in Fig. 5(a) was notch RMG, where the muscle motion was coupled to an RF coaxial cable with an open notch leaking out a portion of the EM energy [45]. The notch RMG has the potential to miniaturize the sensor size, and can be adapted to flexible wearables. The second design in Fig. 5(b) is a non-contact square box with the antenna sensors attached to the inside walls. Forearm can be placed into the box freely without direct contact. The third design in Fig. 5(c) is by the same sensing antenna, but placed on the wristband, which can be convenient for integration into the smart watch as a new input method. Present user interface by fingers on the smart watch display has been impeded by the small screen size, and hand gestures can be a promising alternative [45]. Table II presents the HGR accuracies using the above three designs. For three design variations, notch RMG showed the highest accuracy, which can be favorable in certain applications. Box RMG can still attain reasonable accuracy by the non-contact setup, which further enhance the design flexibility over clothing or in armrests. Wrist RMG showed lower accuracy than the forearm placement because tendon motion had less dielectric contrast than the muscle motion.

To validate that strong RF coupling was from the forearm muscles and not from the direct hand motion in the radar mode, we conducted measurements with the hand inside a radar-absorption-material (RAM) box, as shown in Fig. 5(d), where minimal difference in collected waveforms and achievable HGR accuracy was observed.

## VI. BENCHMARK WITH SEMG

In this section, we performed RMG with synchronous sEMG for the baseline comparison and physiological correlation. The two sensing schemes can be complementary in operation to establish the complete physiological sequence of stimulation and actuation, as well as to study the neuromuscular disorders.

### A. RMG and sEMG placement

For the reference sEMG setup, we used BIOPAC MP36R with the SS2LB leads set and EL503 electrodes (BIOPAC Systems, Goleta, CA). Fig. 5(e) shows the experimental setup with RMG and EMG both on the forearm. Each EMG channels has 3 electrodes on skin as +, −, and ground. We used 2 sEMG channels on the anterior and posterior sides of the forearm. The ground electrodes for two sEMG channels were both placed close to a wrist spot with minimal muscles. RMG and sEMG channels were synchronized on Labview. We performed the same study protocol on two participants as Exp1 and Exp2 in Table III.. The two participants had the same sEMG placement, where + and − electrodes were on the two longitudinal sides of the RMG armband to capture more differential signals with a large distance. Exp3 was the same participant as Exp1, but had a different sEMG placement where the + and − electrodes were on the lower position from the RMG armband. The smaller distance would measure only the muscles close to the electrodes with less voltage resolution.

### B. RMG and sEMG waveform comparison

TABLE III. ACCURACY COMPARISON OF RMG VS. SEMG

|  | Exp1 | Exp2 | Exp3 | Mean |
|---|---|---|---|---|
| RMG | 99.0% | 98.5% | 98.7% | 98.7% |
| sEMG | 68.2% | 70.8% | 66.7% | 68.6% |

As our sEMG waveforms were noisy during the hand gestures, we added two pre-processing procedures: Enveloping the raw data by spline interpolation over local maxima, and smoothing by moving average [46]. The subsequent signal transformation and learning models were the same for sEMG and RMG. The overall HGR accuracy by 7-fold CV is shown in Table III. Accuracy of sEMG was relatively low in comparison with RMG in our setup, which may be caused by the small number of sEMG channels under the large number of gesture classes. Our sEMG implementation was mainly for comparative purposes and was far from ideal. A more comprehensive comparison with the literature results will be presented in next section.

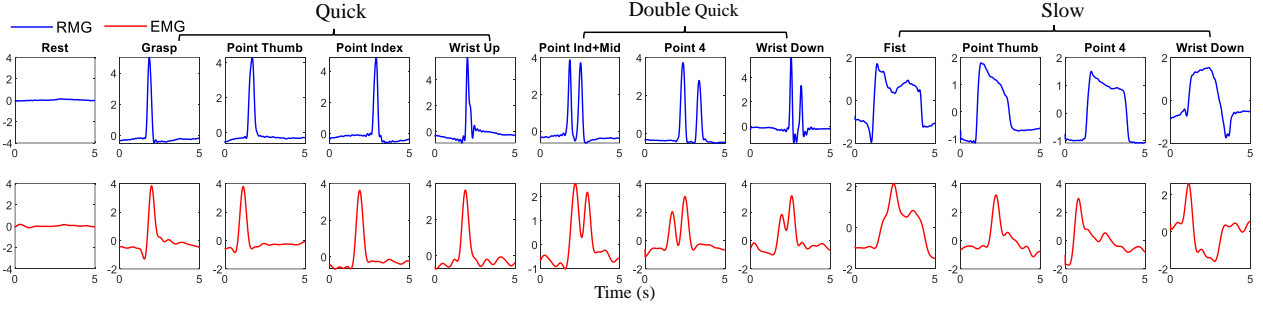

**Fig. 6.** RMG and sEMG waveforms for various gestures by DTW averaging on all samples with the same gesture.

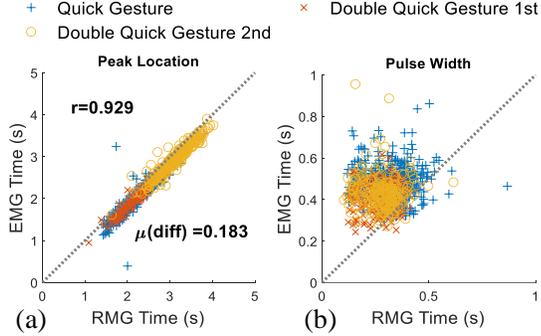

**Fig. 7.** Scatter plots of RMG and sEMG for peak location and pulse width during quick gestures.

As shown in Fig. 6, we also compared the averaged waveforms of different gestures obtained from RMG and sEMG using global dynamic time warping (DTW) [47]. Each gesture had a time window of 5 s, while the y-axis was the normalized amplitude. The RMG waveforms were examples from Tx2-Rx2 and the sEMG from channel 2, both positioned on the posterior side of the forearm. For quick and double quick gestures, both RMG and sEMG presented sharp peaks corresponding to the fast muscle motion. However, compared with RMG, sEMG signals had longer duration of pulse waveforms and showed more tailing after the gesture motion terminated. For slow gestures, RMG showed a more consistent square-wave pattern from the holding period. The sEMG signal showed a shorter pulse duration for gestures that do not require continuous myoelectrical simulation such as the point-finger gestures. For other gestures that require continuous efforts to maintain the position such as the wrist up/down, the sEMG pulse duration were extended. During 'Rest' and between gestures with no intended hand motion, sEMG had more interference and ambiguity due to either hardware sources such as inconsistent electrode contact resistance or from biological sources such as the neural signals from vital signs [48]. In comparison, RMG is less susceptible to vital signs or noises from electrode contacts.

To compare the waveform features further, we performed peak detection in the 14 quick gestures. Fig. 7(a) shows the scatter plot of peak locations of quick gestures in synchronous RMG and sEMG in all samples, where the Pearson correlation coefficient $r = 0.929$ and the mean time difference is a delay of 0.183 s, i.e., RMG and sEMG have a high temporal correlation and a consistent time lag. This delay may indicate the time offset between neural stimulation and muscle actuation. Fig. 7(b) compares the feature of the pulse width, computed as the time duration between the points to the left and right of the half peak magnitude. Most data points are scattered above $y = x$ line, which indicates that RMG waveforms have sharper peaks with less spreading during the quick gestures. Note that the few outliers are probably due to peak detection errors caused by the cases of questionable signal quality. For slow gestures, peak detection is not an appropriate comparison because the waveform features are not always consistent, especially for sEMG.

*C. Timing and latency of RMG*

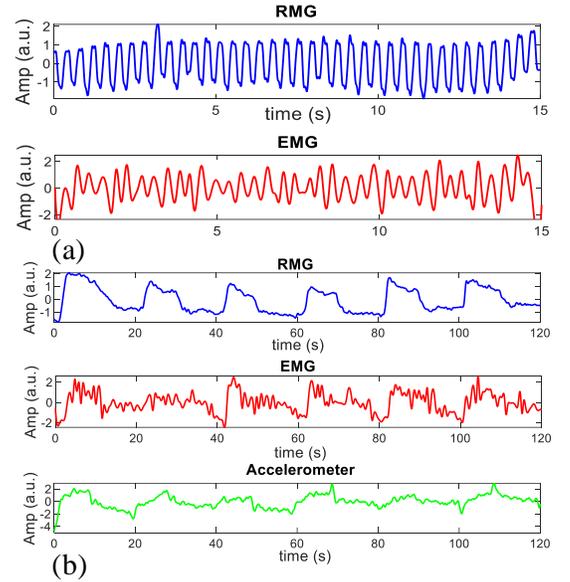

**Fig. 8.** Waveforms recorded from RMG, sEMG, and accelerometer for (a) fast finger motion; (b) slow grasps in 3 times/min with equal strength.

RMG has ultra-low latency with the sampling rate readily over 100,000 samples per second (Sps), which is important for dynamic HGR. Here, we performed the high-speed gesture tracking by RMG and sEMG. The participant followed a metronome of 150 beats/minute and performed the gesture of 'point index and middle fingers' with equal strength at each beat. The sensor setup was the same as Fig. 5(d). The waveforms from one of each RMG and sEMG channels are shown in Fig. 8(a). We can observe from the time waveforms that RMG has a consistent signal pattern corresponding to the quick motions, while sEMG has more fluctuations. Supplementary Table II also presents the statistics of estimated gesture frequency, which is very close to the ground truth with small standard deviation.

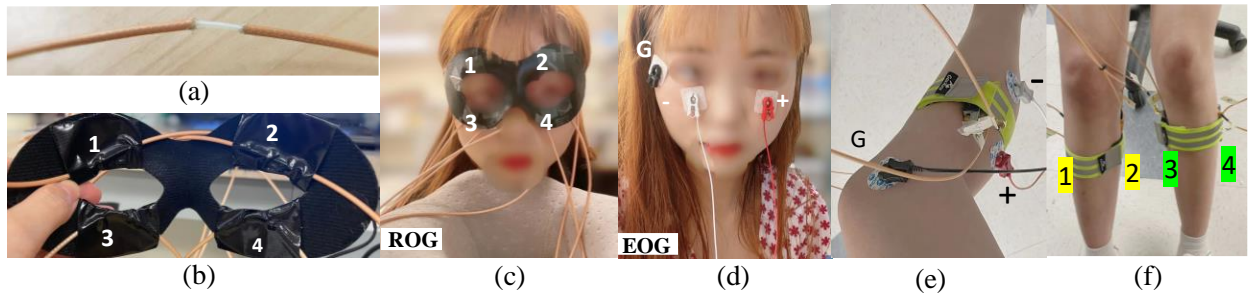

**Fig. 11.** Setup of the ROG and leg RMG systems. (a) One ROG sensor unit by a notched transmission line; (b) Four ROG sensor units on a mask; (c) ROG on a participant's face; (d) EOG setup for baseline comparison; (e) One lower leg RMG sensor unit by whip antennas together with one EMG; (f) Four leg RMG sensor units on two legs.

Compared with surface-motion based sensors including MMG and accelerometers, RMG possesses the unique capability to capture deep muscle contraction. To further corroborate this claim, we tested a slow grip strength detection by RMG together with accelerometers and sEMG. As shown in Table I, during the grip motion, the main muscles include the flexor digitorum superficialis (intermediate), flexor digitorum profondus (deep) and the flexor policus longus (deep) [49]. Hence, the grip motion is mostly by intermediate and deep layers of forearm muscles, instead of superficial layers, which is more difficult for surface motion-based sensors. The participant performed firm holds on the hand dynamometer with a speed of 3 times/minute in equal strength as shown in Fig. 5(f). The waveforms from one of each RMG, sEMG, and accelerometer channels are shown in Fig. 8(b). RMG had a clear and stable signal pattern reflecting the strong and slow grip motion, while sEMG showed some ambiguity and the accelerometer presented even more noisy patterns. This is likely due to the different coupling strength to the deep muscle groups by different sensors. Supplementary Table II also shows the statistics of the estimated grip frequency, where RMG is accurate, and sEMG and accelerometer have more errors. The slow grip frequency is estimated by counting the number of detected cycles each including pinching the hand then back to relaxation according to the voice instruction.

*D. Extension to eye and leg RMG*

To validate the general applicability of RMG to different skeletal muscles, we further extended the setup to wearable radiooculogram (ROG) on eyes and RMG on legs.

As shown in Fig. 9(a), the ROG system integrated four notch RMG to a facemask as shown in Fig. 9(b). A participant wearing ROG and EOG was shown in Figs. 9(c)(d). In a human study of 5 subjects, participants were instructed to move eyes in four directions (Up, Down, Left, and Right) with eyes closed, all of which had 2 versions of moving once and twice. Hence, we had 8 distinctive eye movements (EM), and each motion was performed in a time window of $T_{win}$ = 5 s with around 24 repetitions for every participant (Supplementary video). Then the training model within each participant was built and 7-fold CV was performed to estimate the mean accuracy for each participant. ROG achieved an overall accuracy of 94.2% (Supplementary Fig. 5(a)). ROG can monitor fine eye muscle activities with eyes open or shut. In the future, ROG can be applied for sleep REM and dream stage monitoring [50], and facilitate HCI applications using eye motion control.

Another extension is for monitoring lower leg muscles. We implemented 2 RMG sensing units on each leg with EMG reference for comparison, as shown in Figs. 9(e)(f). We tested 7 postures: 1) tiptoe standing; 2) tiptoe sitting; 3) reverse tiptoe standing; 4) reverse tiptoe sitting; 5) tiptoe sitting with only the right foot; 6) tiptoe sitting with only the left foot; 7) squat (Supplementary video). Each posture was also performed in a time window of $T_{win}$ = 5 s with around 34 repetitions. Leg RMG achieved accuracy of 100% for one participant using 7-fold CV (Supplementary Fig. 5(b)). RMG on lower legs can monitor body postures and can be applied for balance training and fall warning [51].

## VII. DISCUSSION

*A. Comparison to previous HGR works*

A comparison of RMG to previous HGR systems is presented in Table IV. Many existing systems can only recognize 6-9 gestures, partly due to the insufficient representation of forearm muscles. Some works that had over ten gesture classes still faced limitations in the final accuracy. Savur *et al.* [14] can only achieve accuracy of 79.4% for 27 gestures on a single subject, and only 61.0% on 10 subjects. Côté-Allard *et al.* [42] achieved 69.0% for 18 gestures, and the recent work from Moin *et al.* [27] showed 92.9% for 21 gestures. The increasing number of gestures inevitably posed more challenges for classification. In this work, we demonstrated the competitive RMG to recognize 23 gestures with accuracy up to 99.0%. Moreover, the wearable and armrest RMG setups without requiring direct skin contact or restricting the capture volume offer inherent operational advantages over sEMG and camera-based systems. Our choice of ViT classification on spectrograms also shows better

9TABLE IV. COMPARISON TO PREVIOUS WORKS

|  | Li 2019 [17] | Zhang 2016 [25] | Zhang 2015 [31] | McIntosh 2016 [30] | Savur 2016 [14] | Qi 2020 [29] | Côté-Allard 2019 [42] | Moin 2021 [27] | **This work** |
|---|---|---|---|---|---|---|---|---|---|
| Class | 8 | 8 | 5 | 15 | 27 | 9 | 7/18 | 13/21 | **23** |
| Subject | 5 | 4 | 10 | 12 | 1 | - | 17/10 | 2 | **8** |
| Sensor | Camera | FMCW Radar | EIT | sEMG+ pressure | sEMG | sEMG | sEMG | sEMG | **RMG** |
| Model | CNN | CNN | SVM | SVM | Ensemble | GRNN | ConvNet | Neural | **ViT** |
| Accuracy | 98.5% | 96.0% | 97% (hand) 87% (pinch) | 95.8% | 79.4% | 95.3% | 98.3% (7) 69.0% (18) | 97.1% (13) 92.9% (21) | **99.0%** |

performance than traditional ML models adopted in previous works.

*B. Potential future improvements*

*1. Sensor hardware improvement*

In future hardware implementations, we should be able to miniaturize RMG into convenient and comfortable packages as all-in-one wireless wearables, because the expected power consumption and data bandwidth are both very low in today's RF devices. The notch RMG offers a promising design path to reduce cost, form factors, and complexity, especially for integration with a wristwatch.

*2. Real-time classification for HCI*

For HCI in robotic and gaming control, real-time HGR with minimal latency is an important feature. Embedded learning capability with local signal processing and accurate HGR output of RMG will be attractive to many applications.

*3. Fusion with sEMG*

sEMG can estimate neural stimulation of muscle actuation, and RMG can directly detect the actual muscle change. Instead of viewing RMG as a replacement of sEMG, the two sensors can be combined for a fuller physiological interpretation. Our consistent observation of the RMG delay from sEMG possibly indicated the non-trained muscle actuation without the participation of proprioceptive neurons, which can be promising for neuromuscular disorder diagnosis with RMG and sEMG fusion.

*4. Closed-loop EMS control*

A closed-loop control of EMS is another possible future application. EMS has long been employed to either supplement or substitute voluntary muscle stimulation in many settings of rehabilitation and electroceuticals [52]. However, inadequate EMS due to personal and daily differences can cause confusion of antagonistic and synergistic coordination of the muscle groups, and even induce serious spasm. For a more precise control on EMS, RMG can give feedback on actual muscle actuation to control the EMS signal with higher adaptivity to personal and conditional variations.

VIII. CONCLUSION

We have reported a new muscle monitoring technique, named as radiomyography (RMG), which can directly measures the muscle motion by coupling RF energy to superficial and deep internal muscles. Operation over clothing without direct skin touch enables convenient setup and comfortable operation.

The MIMO approach enriches the collected information with a small number of sensing points. We implemented RMG as a wearable forearm sensor to accurately track forearm muscles activating hand gestures. For the HGR purpose, we adopted the vision transformer as the classification model and effectively boosted the accuracy up to 99.0% for 23 hand gestures tested on 8 participants. We further adopted transfer learning to address cross-subject and operational variations. For HGR systems, RMG has lower cost, lower complexity, lower latency and less privacy issues than camera-based devices, as well as higher user comfort and accuracy than contact-based devices.

RMG has the unique advantage to monitor internal muscles non-invasively. In the future, RMG and EMG can be fused together to derive the full information of stimulation and actuation. RMG can lead to new methods for assessment of muscle functions, monitoring of muscle fatigue, and diagnosis of neuromuscular disorders. RMG is also promising for future HCI applications including exoskeleton robotic control, virtual reality interface, and in-air gesture capture.

REFERENCES

[1] M. Cifrek, V. Medved, S. Tonković, and S. Ostojić, "Surface EMG based muscle fatigue evaluation in biomechanics," *Clin. Biomech.*, vol. 24, no. 4, pp. 327-340, 2009.
[2] D. Leonardis et al., "An EMG-controlled robotic hand exoskeleton for bilateral rehabilitation," *IEEE Trans. Haptics*, vol. 8, no. 2, pp. 140-151, 2015.
[3] P. K. Jamwal, S. Hussain, Y. H. Tsoi, and S. Q. Xie, "Musculoskeletal model for path generation and modification of an ankle rehabilitation robot," *IEEE Trans. Hum. Mach. Sys.,* vol. 50, no. 5, pp. 373-383, 2020.
[4] I. Moon, M. Lee, J. Chu, and M. Mun, "Wearable EMG-based HCI for electric-powered wheelchair users with motor disabilities," in *Proc. IEEE Intl. Conf. Robotics and Automation*, 2005, pp. 2649-2654.
[5] Q. Qian, X. Hu, Q. Lai, S. C. Ng, Y. Zheng, and W. Poon, "Early stroke rehabilitation of the upper limb assisted with an electromyography-driven neuromuscular electrical stimulation-robotic arm," *Front. Neurol., Clinical Trial*, vol. 8, Sept. 2017.
[6] M. B. I. Reaz, M. S. Hussain, and F. Mohd-Yasin, "Techniques of EMG signal analysis: detection, processing, classification and applications," *Biol. Proced.,* vol. 8, no. 1, pp. 11-35, 2006.
[7] M. O. Ibitoye, N. A. Hamzaid, J. M. Zuniga, and A. K. A. Wahab, "Mechanomyography and muscle function assessment: A review of

current state and prospects," *Clin. Biomech.,* vol. 29, no. 6, pp. 691-704, 2014.
[8] C. D. Lee, Y. Song, A. C. Peltier, A. A. Jarquin-Valdivia, and P. D. Donofrio, "Muscle ultrasound quantifies the rate of reduction of muscle thickness in amyotrophic lateral sclerosis," *Muscle Nerve,* vol. 42, no. 5, pp. 814-819, 2010.
[9] E. Mercuri, A. Pichiecchio, J. Allsop, S. Messina, M. Pane, and F. Muntoni, "Muscle MRI in inherited neuromuscular disorders: past, present, and future," *J. Magn. Reson. Imaging,* vol. 25, no. 2, pp. 433-440, 2007.
[10] J. S. Sonkusare, N. B. Chopade, R. Sor, and S. L. Tade, "A review on hand gesture recognition system," in *2015 International Conference on Computing Communication Control and Automation*, 2015, pp. 790-794.
[11] R. Z. Khan and N. A. Ibraheem, "Hand gesture recognition: A literature review," *Int. J. Artif.,* vol. 3, no. 4, p. 161, 2012.
[12] J. DelPreto and D. Rus, "Plug-and-play gesture control using muscle and motion sensors," in *Proc. ACM/IEEE Intl. Conf. Human-Robot Interaction*, 2020, pp. 439-448.
[13] H. Liu and L. Wang, "Gesture recognition for human-robot collaboration: A review," *Int. J. Ind. Ergon.,* vol. 68, pp. 355-367, 2018.
[14] C. Savur and F. Sahin, "American Sign Language Recognition system by using surface EMG signal," in *IEEE Intl. Conf. Systems, Man, and Cybernetics (SMC)*, 2016, pp. 002872-002877.
[15] K. A. Smith, C. Csech, D. Murdoch, and G. Shaker, "Gesture recognition using mm-wave sensor for human-car interface," *IEEE Sens. Lett.*, vol. 2, no. 2, pp. 1-4, 2018.
[16] S. S. Rautaray and A. Agrawal, "Vision based hand gesture recognition for human computer interaction: a survey," *Artif. Intell. Rev.*, vol. 43, no. 1, pp. 1-54, 2015.
[17] G. Li et al., "Hand gesture recognition based on convolution neural network," *Cluster Comput.*, vol. 22, no. 2, pp. 2719-2729, 2019.
[18] R. H. Venkatnarayan and M. Shahzad, "Gesture recognition using ambient light," *Proc. ACM Interact. Mob. Wearable Ubiquitous Technol.*, vol. 2, no. 1, pp. 1-28, 2018.
[19] J. Suarez and R. R. Murphy, "Hand gesture recognition with depth images: A review," in *IEEE RO-MAN: 21st IEEE Intl. Symp. Robot and Human Interactive Communication*, 2012, pp. 411-417.
[20] B. Feng et al., "Depth-projection-map-based bag of contour fragments for robust hand gesture recognition," *IEEE Trans. Hum. Mach. Sys.,* vol. 47, no. 4, pp. 511-523, 2016.
[21] D. Mazumdar, A. K. Talukdar, and K. K. Sarma, "Gloved and free hand tracking based hand gesture recognition," in *1st Intl. Conf. Emerging Trends and Applications in Computer Science*, 2013, pp. 197-202.
[22] C. Xu, P. H. Pathak, and P. Mohapatra, "Finger-writing with smartwatch: A case for finger and hand gesture recognition using smartwatch," in *Proc.16th International Workshop on Mobile Computing Systems and Applications*, 2015, pp. 9-14.
[23] J. Lien et al., "Soli: Ubiquitous gesture sensing with millimeter wave radar," *ACM Trans. Graphics*, vol. 35, no. 4, pp. 1-19, 2016.
[24] X. Gao et al., "Barcode based hand gesture classification using AC coupled quadrature Doppler radar," in *IEEE MTT-S Intl. Microwave Symp. (IMS)*, 2016, pp. 1-4.
[25] Z. Zhang, Z. Tian, and M. Zhou, "Latern: Dynamic continuous hand gesture recognition using FMCW radar sensor," *IEEE Sens. J.*, vol. 18, no. 8, pp. 3278-3289, 2018.
[26] K. S. Krishnan, A. Saha, S. Ramachandran, and S. Kumar, "Recognition of human arm gestures using Myo armband for the game of hand cricket," in *IEEE Intl. Symp. Robotics and Intelligent Sensors (IRIS)*, 2017, pp. 389-394.
[27] A. Moin et al., "A wearable biosensing system with in-sensor adaptive machine learning for hand gesture recognition," *Nat. Electron.*, vol. 4, no. 1, pp. 54-63, 2021.
[28] U. Côté-Allard, C. L. Fall, A. Campeau-Lecours, C. Gosselin, F. Laviolette, and B. Gosselin, "Transfer learning for sEMG hand gestures recognition using convolutional neural networks," *IEEE Intl. Conf. Systems, Man, and Cybernetics (SMC)*, 2017, pp. 1663-1668.
[29] J. Qi, G. Jiang, G. Li, Y. Sun, and B. Tao, "Surface EMG hand gesture recognition system based on PCA and GRNN," *Neural. Comput. Appl.*, vol. 32, no. 10, pp. 6343-6351, 2020.
[30] J. McIntosh, C. McNeill, M. Fraser, F. Kerber, M. Löchtefeld, and A. Krüger, "EMPress: Practical hand gesture classification with wrist-mounted EMG and pressure sensing," in *Proc. CHI Conf. Human Factors in Computing Systems*, 2016, pp. 2332-2342.
[31] Y. Zhang and C. Harrison, "Tomo: Wearable, low-cost electrical impedance tomography for hand gesture recognition," *Proc. 28th Annual ACM Symposium on User Interface Software & Technology*, 2015, pp. 167-173.
[32] X. Hui and E. C. Kan, "Monitoring vital signs over multiplexed radio by near-field coherent sensing," *Nat. Electron.*, vol. 1, no. 1, pp. 74-78, 2018.
[33] X. Hui, T. B. Conroy, and E. C. Kan, "Multi-point near-field RF sensing of blood pressures and heartbeat dynamics," *IEEE Access*, vol. 8, pp. 89935-89945, 2020.
[34] A. Dosovitskiy et al., "An image is worth 16x16 words: Transformers for image recognition at scale," *arXiv preprint* arXiv:2010.11929, 2020.
[35] Y. Kim and B. Toomajian, "Application of Doppler radar for the recognition of hand gestures using optimized deep convolutional neural networks," *11th European Conf. Antennas and Propagation (EUCAP)*, 2017, pp. 1258-1260.
[36] X. Hui, P. Sharma, and E. C. Kan, "Microwave stethoscope for heart sound by near-field coherent sensing," *IEEE MTT-S Intl. Microwave Symp. (IMS)*, pages 365-368. IEEE, 2019.
[37] X. Hui and E. C. Kan, "Seat integration of RF vital-sign monitoring," *IEEE MTT-S International Microwave Biomedical Conference (IMBioC)*, pages 1-3, 2019.
[38] Z. Zhang, P. Sharma, T. B. Conroy, V. Phongtankuel, and E. C. Kan, "Objective scoring of physiologically induced dyspnea by non-invasive RF sensors," *IEEE. Trans. Biomed. Engr.*, vol. 69, no. 1, pp. 432-442, 2021.
[39] P. Sharma, X. Hui, J. Zhou, T. B. Conroy, and E. C. Kan, "Wearable radio-frequency sensing of respiratory rate, respiratory volume, and heart rate," *NPJ Digit. Med.*, vol. 3, p. 98, July 2020.
[40] J. Zhou, P. Sharma, X. Hui, and E. C. Kan, "A wireless wearable RF sensor for brumation study of chelonians," *IEEE J. Electromagn. RF Microw. in Medicine and Biology (J-ERM),* vol. 5, no. 1, pp. 17-24, 2020.
[41] R. E. Fields, "Evaluating compliance with FCC guidelines for human exposure to radiofrequency electromagnetic fields," *OET bulletin,* vol. *65, no.* 10, 1997.
[42] U. Côté-Allard et al., "Deep learning for electromyographic hand gesture signal classification using transfer learning," *IEEE Trans. Neural Syst.*, vol. 27, no. 4, pp. 760-771, 2019.
[43] A. Vaswani *et al.*, "Attention is all you need," *Adv. Neural. Inf.,* vol. 30, 2017.
[44] C. Tan, F. Sun, T. Kong, W. Zhang, C. Yang, and C. Liu, "A survey on deep transfer learning," in *Intl. Conf. Artificial Neural Networks*, 2018, pp. 270-279.
[45] Z. Zhang, P. Sharma, J. Zhou, X. Hui, and E. C. Kan, "Furniture-integrated respiration sensors by notched transmission lines," *IEEE Sens. J.*, vol. 21, no. 4, pp. 5303-5311, Feb. 2021.
[46] A. Rahimi, S. Benatti, P. Kanerva, L. Benini, and J. M. Rabaey, "Hyperdimensional biosignal processing: A case study for EMG-based hand gesture recognition," in *IEEE Intl. Conf. Rebooting Computing (ICRC)*, 2016, pp. 1-8.
[47] F. Petitjean, A. Ketterlin, and P. Gançarski, "A global averaging method for dynamic time warping, with applications to clustering," *Pattern recognition,* vol. 44, no. 3, pp. 678-693, 2011.
[48] N. Amrutha and V. Arul, "A review on noises in EMG signal and its removal," *Int. J. Sci. Res. Publ,* vol. 7, no. 5, pp. 23-27, 2017.
[49] R. W. Bohannon, "Muscle strength: Clinical and prognostic value of hand-grip dynamometry," *Curr. Opin. Clin. Nutr. Metab. Care,* vol. 18, no. 5, pp. 465-470, 2015.
[50] N. Cooray, F. Andreotti, C. Lo, M. Symmonds, M. T. Hu, and M. De Vos, "Detection of REM sleep behaviour disorder by automated polysomnography analysis," *Clin. Neurophysiol.,* vol. 130, no. 4, pp. 505-514, 2019.
[51] T. Isezaki *et al.*, "Sock-type wearable sensor for estimating lower leg muscle activity using distal EMG signals," *Sensors,* vol. 19, no. 8, p. 1954, 2019.
[52] S. Zhang, X. Zhang, S. Cao, X. Gao, X. Chen, and P. Zhou, "Myoelectric pattern recognition based on muscle synergies for simultaneous control of dexterous finger movements," *IEEE Trans. Hum. Mach. Sys.,* vol. 47, no. 4, pp. 576-582, 2017.